# When do composite estimands answer non-causal questions?

Brennan C Kahan[1], Tra My Pham[1], Conor Tweed[1], Tim P Morris[1]

[1] MRC Clinical Trials Unit at UCL, London, UK

Correspondence to: Brennan Kahan (b.kahan@ucl.ac.uk)

**Abstract**

Under a composite estimand strategy, the occurrence of the intercurrent event is incorporated into the endpoint definition, for instance by assigning a poor outcome value to patients who experience the event. Composite strategies are sometimes used for intercurrent events that result in changes to assigned treatment, such as treatment discontinuation or use of rescue medication. Here, we show that a composite strategy for these types of intercurrent events can lead to the outcome being defined differently between treatment arms, resulting in estimands that are not based on causal comparisons. This occurs when the intercurrent event can be categorised, such as based on its timing, and at least one category applies to one treatment arm only. For example, in a trial comparing a 6 vs. 12-month treatment regimen on an "unfavourable" outcome, treatment discontinuation can be categorised as occurring between 0-6 or 6-12 months. A composite strategy then results in treatment discontinuations between 6-12 months being part of the outcome definition in the 12-month arm, but not the 6-month arm. Using a simulation study, we show that this can dramatically affect conclusions; for instance, in a scenario where the intervention had no direct effect on either a clinical outcome or occurrence of the intercurrent event, a composite strategy led to an average risk difference of –10% and rejected the null hypothesis almost 90% of the time. We conclude that a composite strategy should not be used if it results in different outcome definitions being used across treatment arms.

**Key words:** estimand, composite strategy, causal effect, ICH E9(R1), composite outcome

**Background**

The use of estimands to help describe research questions has become increasingly popular in randomised trials [1-7]. The ICH-E9(R1) addendum provides a standard structure for defining estimands, with one of the most important aspects being which strategies will be used to address intercurrent events [1]. An intercurrent event is a post-randomisation event which affects either the interpretation or existence of the outcome [1, 6]. Common intercurrents events are treatment discontinuation, treatment switching, failure to start assigned treatment, and death [8-10].

A commonly used strategy to address intercurrent events is a composite strategy [8-10]. Under a composite strategy, the intercurrent event is incorporated into the outcome definition [1, 6]. For instance, for a symptom score outcome, patients with the intercurrent event could be assigned a high value of the score (i.e. severely symptomatic), or, if the outcome is a clinical event such as disease recurrence, the outcome could be modified to be either disease recurrence or occurrence of the intercurrent event. While the composite strategy is often used to handle death when it acts as an intercurrent event, it is also sometimes used to handle intercurrent events related to changes in assigned treatment, such as treatment discontinuation, treatment switching, or use of rescue therapy [10]. For example, many trials in tuberculosis or HIV use an "unfavourable" or "treatment failure" outcome, which includes treatment discontinuation or treatment switching being defined as unfavourable/a failure [11-14].

A perceived benefit of the composite strategy over other strategies, such as the hypothetical or principal stratum strategy, may be that estimation of the composite strategy preserves randomisation and thus is not subject to potential bias. In contrast, estimation of principal stratum and hypothetical strategies typically rely on strong, untestable assumptions, and may be subject to bias when these assumptions are not fulfilled [6, 15-23].

However, the use of a composite strategy can still sometimes lead to a non-causal question being addressed [6]. This is not due to bias, which is defined as a systematic deviation of the estimator from the estimand's true value, but rather due to a problem with the way the composite estimand is defined, which means it can no longer be viewed as a proper causal effect. This occurs when the composite strategy leads to the outcome being defined differently between the treatment arms. This can happen when the intercurrent event itself is defined differently between treatment arms, due to inherent differences between the treatments being assessed.

For example, in a trial comparing surgery vs. a wait-and-see approach on being pain free at 12 months in individuals with a knee injury, the intercurrent event of "unsuccessful surgery" (where the procedure cannot be completed), the intercurrent event applies to the surgery in the treatment arm only. Therefore, using a composite strategy would lead to the outcome being defined as: (i) being pain free, for individuals in the wait-and-see treatment group; and (ii) being pain free *and* not having an unsuccessful surgery, for individuals in the surgery group. Because the outcome is defined differently between treatment arms, the between-arm comparison cannot be viewed as a valid causal effect, even though it is based on a randomised comparison [6]. Other examples of how a composite strategy can lead to the outcome being defined differently between treatment arms are shown in Table 1.

The purpose of this article is to describe when composite estimands lead to non-causal comparisons, and to demonstrate to what extent this can affect interpretation of results. We provide several examples of trials where this could occur (Table 1) and show mathematically when a composite strategy can create artificial differences between treatment groups by leading to non-comparable outcome definitions between arms. We then evaluate to what extent this can affect conclusions in a small simulation study. A summary of key messages is provided in Box 1.

**How composite estimands can lead to non-causal comparisons**

A key concept of a causal comparison is to compare the *same* outcome between different treatments conditions, to determine how that outcome changes between the treatments. If the outcome is defined different across treatments, then the comparison would not generally be a causal one [24]. We illustrate below how the use of a composite strategy for intercurrent events such as treatment discontinuation can lead to outcomes being defined differently between treatment arms, and hence a non-causal comparison. For illustration purposes, we use a tuberculosis trial comparing two different treatment durations (6 months vs. 12 months) on an "unfavourable" outcome (e.g. not free from tuberculosis at the end of 12 months). The intercurrent event is treatment discontinuation (i.e. stopping assigned treatment early). We show below how using a composite strategy in this example leads to a non-causal comparison.

Let $Z$ denotes treatment assignment, with $Z = 1$ denoting a 6-month strategy, and $Z = 0$ a 12-month strategy. An outcome, $Y$, is defined as the "unfavourable" outcome (where $Y = 1$ denotes an unfavourable outcome, and $Y = 0$ denotes a favourable one). $Y$ is defined as a composite of $Y_a$ and $Y_b$, where $Y_a$ denotes a patient's tuberculosis status at 12 months (1=positive, 0=negative), and $Y_b$ denotes occurrence of the intercurrent event (1=discontinued treatment early, 0=did not discontinue treatment early). Thus, $Y = 1$ if either $Y_a$ or $Y_b = 1$.

Here, the component $Y_b$ is defined *differently* between the treatment conditions. For clarity, we further separate $Y_b$ into two components, $Y_{b,0-6}$ and $Y_{b,6-12}$, where $Y_{b,0-6}$ denotes whether a patient discontinued treatment between 0-6 months, and $Y_{b,6-12}$ denotes whether they discontinued between 6-12 months. Of note, $Y_{b,0-6}$ can be either 0 or 1 in both treatment arms, however $Y_{b,6-12}$ by definition can only be 0 for the 6-month group ($Z = 1$), but can take values 0 or 1 in the 12-month group ($Z = 0$).

Thus, the overall outcome $Y$ is defined as follows in the 12-month group:

$$Y|Z = 0 = \begin{cases} 0 \text{ if } Y_a = 0, \quad Y_{b,0-6} = 0, \quad Y_{b,6-12} = 0 \\ 1 \text{ otherwise} \end{cases}$$

Conversely, in the 6-month arm, the outcome $Y$ is defined as:

$$Y|Z = 1 = \begin{cases} 0 \text{ if } Y_a = 0, \quad Y_{b,0-6} = 0 \\ 1 \text{ otherwise} \end{cases}$$

This leads to the very odd situation shown in Figure 1, where a patient who received 6 months of treatment and was alive and TB-free at 12-months would be considered a *success* in the 6-month group, but a *failure* in the 12-month group, despite the fact they have identical data under both treatments. Similarly, a patient could be considered a failure in the 12-month group but not the 6-month group due to treatment discontinuation, even if they receive *more* treatment in the 12-month arm and would have a good clinical outcome under either treatment.

This use of the composite strategy to define the outcome leads to the following non-causal comparison being made between treatment conditions:

$$E(Y^1 - Y^0) = E(Y_a^1 - Y_a^0) + E(Y_{b,0-6}^1 - Y_{b,0-6}^0) - E(Y_{b,6-12}^0)$$

where $Y^1$ and $Y^0$ denote potential outcomes [25, 26] under assignment to the 6- and 12-month treatments respectively, and $Y_a^1$, $Y_a^0$, $Y_{b,0-6}^1$, $Y_{b,0-6}^0$, and $Y_{b,6-12}^0$ similarly denote potential outcomes for the components. From the above expression, it is easy to see that $E(Y_a^1 - Y_a^0)$ and $E(Y_{b,0-6}^1 - Y_{b,0-6}^0)$ are both causal contrasts, however the final term in the expression, $E(Y_{b,6-12}^0)$, is not. This is because it pertains to the control condition only, with no corresponding outcome defined under the intervention. As such, this expression will produce a non-zero treatment effect when $E(Y_{b,6-12}^0) > 0$ (i.e. when some patients discontinue treatment between 6-12 months under the control condition), even when $E(Y_a^1 - Y_a^0) = 0$ and $E(Y_{b,0-6}^1 - Y_{b,0-6}^0) = 0$ (i.e. when treatment has no direct effect on either the clinical outcome or the occurrence of the intercurrent event).

Of note, the expression above assumes that the different components are mutually exclusive, which is not the case. However, expressing it in this way helps to clarify the fundamental issue in this comparison, which is the presence of $E(Y_{b,6-1}^0)$.

**Identifying when a composite strategy will lead to non-causal estimands**

A simple principle can be used to identify whether the use of a composite strategy will lead to a non-causal estimand: if the intercurrent event can be categorised such that at least one category applies to one treatment condition only, then a composite strategy to handle the intercurrent event will lead to the outcome being defined different between treatments, which in turn will lead to a non-causal estimand. For instance, in the tuberculosis example used in the previous section, the intercurrent event can be categorised into two categories ($Y_{b,0-6}$ and $Y_{b,6-12}$), and one of these categories ($Y_{b,6-1}$) applies only to the control arm. Additional examples are shown in Table 1.

**Simulation study - methods**

We conducted a small proof-of-principle simulation study [27] to demonstrate that a composite strategy can lead to non-causal comparisons when it results in the outcome being defined differently between treatment conditions. The simulation study is motivated by the tuberculosis trial setting described earlier, where two different durations of treatment (6 vs. 12 months) are compared on a clinical outcome (not being free from tuberculosis at 12 months), where a composite strategy is employed for the intercurrent event of stopping treatment early (prior to 6 months in the 6-month group, and prior to 12 months in the 12-month group). These trials would typically use a non-inferiority design, however we use a superiority design here for simplicity.

This simulation study is not intended to be realistic but rather to demonstrate that a composite strategy in this setting can be seriously misleading. To this end, we consider two situations:

*Scenario 1*: treatment duration has no impact on clinical outcomes.
*Scenario 2*: treatment duration of 6 months leads to worse clinical outcomes.

The data-generating mechanism for a single simulation run is as follows. The sample size of the trial is set to $n = 1,000$ participants, with treatment $Z$ being block-randomized so that 500 participants are assigned to $Z = 1$ (6-month treatment duration) and 500 to $Z = 0$ (12-month treatment duration).

The outcomes are defined as:
- $Y_a$ denotes tuberculosis status at 12 months, with 1=not free from tuberculosis and 0=free from tuberculosis
- $Y_b$ denotes occurrence of the intercurrent event, with 1=stopped treatment early (prior to 6 months in the 6-month arm, or prior to 12 months in the 12-month arm) and 0=did not stop treatment early
- $Y$ denotes the composite "unfavourable" outcome and is equal to 1 if $Y_a = 1$ or $Y_b = 1$, and 0 otherwise.

In scenario 1, $Y_a$ (tuberculosis status at 12 months) is identical in each arm and is simulated as $Y_a \sim Bernoulli(0.4)$. We generated treatment discontinuation, $Y_b$, to occur either in months 1–6 or 7–12, with equal probability in each period and with overall probability in the 12-month group of 0.3. This implies that in the 6-month group, the probability of discontinuing early is 0.15. We generated discontinuation in this way so that the probability of early discontinuation was identical in both treatment arms over the first 6 months in which patients in both arms are meant to receive treatment, but greater overall in the 12-month group due to the longer treatment period.

In scenario 2, $Y_a$ (tuberculosis status at 12 months) is different between the two arms. We generated $Y_a \sim Bernoulli(0.4)$ in the 12-month group, and $Y_a \sim Bernoulli(0.5)$ in the 6-month group, fixing it so that the excess adverse clinical outcomes in the 6-month arm occurs only after the first 6 months, i.e. after treatment has been stopped. The intercurrent event status, $Y_b$, was generated the same way as in scenario 1.

The estimand of interest is the risk-difference for the composite "unfavourable" outcome $Y$. We note that, as described earlier, this estimand is not a well-defined causal contrast; as stated above, our aim is to demonstrate the problems associated with using this estimand.

The estimator for this estimand compares the difference in the mean of the composite "unfavourable" outcome in each arm:
$$\hat{\theta} = (\bar{Y}|Z = 1) - (\bar{Y}|Z = 0)$$

Due to the simple data structure (e.g. 1:1 randomisation with a large sample size), this will be estimated using a t-test, as this will return identical point estimates to other models (e.g. an identity-link binomial model or a logistic regression followed by standardisation, if estimated using maximum likelihood). Due to the 1:1 randomisation and large sample size, standard errors will also be valid [28].

Performance under each scenario will be measured using the expected risk difference, $\mathbf{E}(\hat{\theta})$, *rejection fraction* (the proportion of times $H0: \theta = 0$ is rejected using $\alpha = 0.05$, i.e. $\Pr(p < 0.05)$), and the expected number of excess composite outcomes that could only occur in the 12-month arm.

All code was written in Stata, run in version 18 with updates to 14feb2024. No community-contributed packages were used. $n_{reps} = 10,000$ repetitions were run to obtain Monte Carlo standard error (MCSE) of the mean of $\hat{\theta}$ below 0.1% (estimated after an initial run-in of $n_{reps} = 500$).

**Simulation study - results**

Results are shown in Table 2. In scenario 1, where treatment duration has no direct effect on clinical outcomes or on the intercurrent event (apart from there being greater opportunity to discontinue early in the 12-month arm due to the longer time-period), the use of a composite strategy led to a large apparent benefit for the 6-month arm. The mean risk difference was -10 percentage points in favour of the 6-month duration, and the null was rejected 89% of the time. This was despite the fact that both $Y_a$ (tuberculosis status) and $Y_b$ (treatment discontinuation) follow an identical distribution over time in the two treatment arms.

In scenario 2, where the 6-month treatment duration leads to worse clinical outcomes but does not affect the intercurrent event (apart from the longer time-period in the 12-month arm), the use of a composite strategy showed a small benefit for the 6-month arm (mean risk difference of -1.7 percentage points), despite the 6-month arm leading to substantially worse clinical outcomes.

The reason the 6-month arm appeared beneficial despite having no impact on intercurrent event probabilities and either no effect or a harmful effect on the clinical outcome, is because the different ways of defining outcomes between treatment arms lead to an excess of intercurrent events in the 12-month arm (i.e. events that could only occur in the 12-month arm due to the way outcomes were defined). This can be seen in the 'excess events' performance measure (Table 2): an average of 50 extra intercurrent events ($Y_b$) are counted among the 500 participants in the 12-month group in each scenario, and these events occur between months 6 and 12 (i.e. are events that could only occur in the 12-month arm). It is these 'excess' events that are responsible for apparent beneficial effects of the 6-month arm when the 6-month arm either has no effect, or an adverse effect, on the outcome components.

**Discussion**

In this article we have described how a composite strategy can answer non-causal questions. This occurs when the composite strategy leads to outcomes being defined differently between treatment arms. In our simulation study, we found that the problems introduced from this approach can dramatically affect conclusions, for instance by creating an apparent benefit when no such benefit exists, or by masking a harmful treatment.

These findings have clear implications for practice. First, a composite strategy should not be adopted when it results in outcomes being defined differently between treatment arms. Second, when a composite strategy is used for treatment-related intercurrent events, such as treatment discontinuation, justification should be provided to show that it will not lead to disparate outcome definitions across treatment arms. Third, it should be clearly stated in the estimand definition that a composite strategy has been used. This is often described as part of the endpoint definition, rather than explicitly as an intercurrent event strategy, however this approach may not alert readers to potential issues to be aware of. Finally, when a composite strategy is used, each component should be defined as a secondary outcome, in line with current recommendations for composite outcomes [29-33]. A summary of these recommendations is given in Box 2.

Importantly, the issues described here around the composite strategy can also apply to the while-on-treatment strategy. Under this strategy, outcomes prior to the intercurrent event are of interest (e.g. "unfavourable" outcome at 12 months or at the point of treatment discontinuation, whichever is first). This strategy modifies the outcome definition, by changing the timeframe of the outcome. Therefore, similar issues can arise around a while-on-treatment strategy leading to different outcome definitions between treatment groups, when the intercurrent event can be categorised such that some categories apply to one treatment arm only. For instance, if comparing two treatment durations, such as 6-months vs. 12-months, treatment discontinuation between 6 and 12 months apply to the 12-month treatment arm only; thus, the use of a while-on-treatment strategy would lead to an outcome definition of "unfavourable" outcome at 12 months or at the point of treatment discontinuation within 6 months (in the 6-month arm), or "unfavourable" outcome at 12 months or at the point of treatment discontinuation within 12 months (12-month arm).

Despite the potential problems associated with using a composite strategy for treatment-related intercurrent events, such as treatment discontinuation, this approach can be valid so long as its use does not result in different outcome definitions between treatment groups. For instance, in a trial comparing two daily medications for pain relief, where either treatment group could receive rescue medication, there is no categorisation of the intercurrent event (rescue medication) that applies to one treatment arm only. Therefore, the composite strategy would not lead to differential outcome definitions between treatment arms, and so would return a valid causal comparison.

**Conclusion**

The use of a composite strategy to handle intercurrent events can lead to non-causal comparisons when it results in different outcome definitions being used between treatment groups. Therefore, caution should be used when considering this strategy for treatment-related intercurrent events, such as treatment discontinuation.

**Figure 1 – Demonstration of how a composite strategy results in different outcome definitions between treatment arms in a trial comparing 6- and 12-month treatment durations.** The figure shows outcomes for four patients (labelled 1 to 4); for each patient, the figure shows two potential outcomes, one if the patient had been assigned the 6-month duration, and one if they had been assigned the 12-month duration. The top and bottom lines for each patient correspond to the 6- and 12-month durations respectively. In this example, the outcome is a composite of a positive clinical outcome (being free from tuberculosis at 12-months) and occurrence of the intercurrent event (treatment discontinuation; defined as stopping treatment <6 months in the 6-month treatment group, and stopping treatment <12 months in the 12-month treatment group); patients are classified as having a "favourable" outcome if they are free from tuberculosis at 12-months, and do not experience the intercurrent event, and as having an "unfavourable" otherwise A blue line indicates the patient has been classified as "favourable", and a red line denotes "unfavourable". All four patients in the figure have a positive clinical outcome (i.e. are free of tuberculosis at 12 months), but some stop their treatment early (treatment discontinuation).

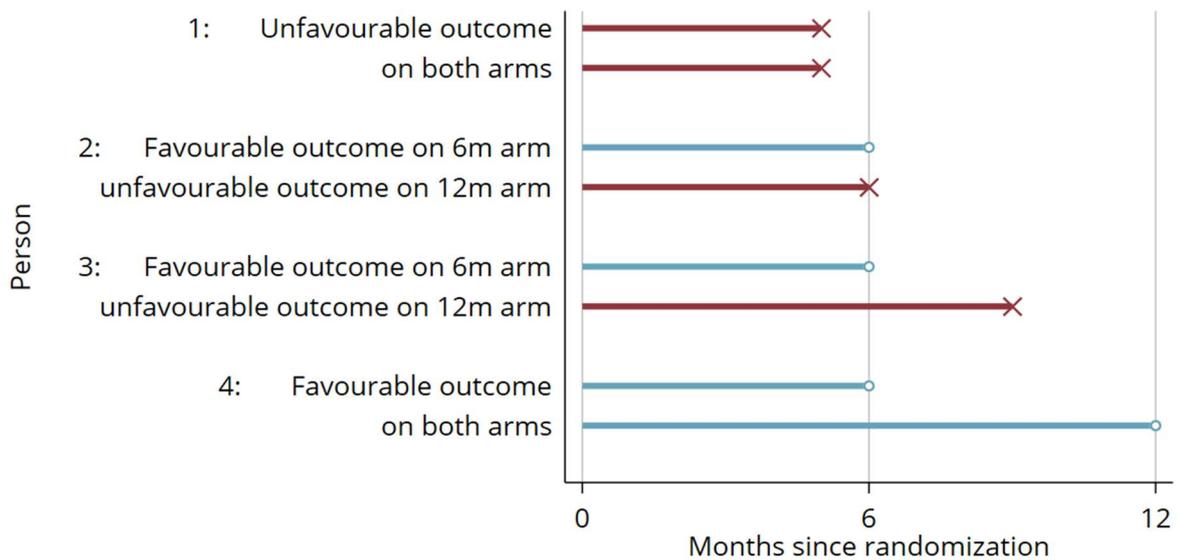

**Table 1 – Examples of when a composite strategy leads to different outcome definitions between treatment arms.**

| Example | Explanation |
|---|---|
| A trial in patients with tuberculosis which compares a 6 month vs. 12 months treatment duration on an "unfavourable" outcome (not being free from tuberculosis at 12 months), where some patients stop their assigned treatment early (the intercurrent event). | The intercurrent event can be categorised into two mutually exclusive categories (i) treatment discontinuation between 0-6 months; and (ii) treatment discontinuation between 6-12 months.<br><br>Category (i) applies to both treatment group, however category (ii) applies only to the 12-month group.<br><br>Therefore, a composite strategy will lead to the following different outcome definitions in each group:<br>-*6-month group*: not free from tuberculosis at 12 months, or stopping treatment before 6 months<br>-*12-month group*: not free from tuberculosis at 12 months, or stopping treatment before 12 months |
| A trial in patients with prostate cancer which compares a new experimental arm vs. standard of care on overall survival at 5 years, where some patients switch from standard of care to the new experimental arm on disease progression (the intercurrent event). | The intercurrent event applies to one arm only (the standard of care arm).<br><br>Therefore, a composite strategy will lead to the following different outcome definitions in each group:<br>-*experimental arm*: alive at 5 years<br>-*standard of care arm*: alive at 5 years without switching to the experimental arm on progression |
| A trial in patients with acute upper gastrointestinal bleeding which compares a restrictive red blood cell transfusion strategy (haemoglobin < 8 g/dL) vs. liberal strategy (haemoglobin < 10 g/dL) on further bleeding within 30 days, where some patients are transfused when their haemoglobin count is above the assigned threshold (the intercurrent event). | The intercurrent event can be categorised into two mutually exclusive categories (i) transfusing between 8-10 g/dL; and (ii) transfusing above 10 g/dL.<br><br>Category (i) applies only to the restrictive transfusion group, while category (ii) applies to both groups.<br><br>Therefore, a composite strategy will lead to the following different outcome definitions in each group:<br>-*restrictive group*: further bleeding within 30 days, or transfusion above 8 g/dL |
| | -*liberal group*: further bleeding within 30 days, or transfusion above 10 g/dL |
| A trial in patients with severe anxiety which compares anxiety medication plus cognitive behavioural therapy (CBT) vs. anxiety medication alone on being free from anxiety at 6 months, where some patients stop taking their assigned treatment, patients in the medication plus CBT group stopping either medication or CBT, and patients in the medication alone group stopping medication (the intercurrent event). | The intercurrent event can be categorised into two mutually exclusive categories (i) stopping anxiety medication early; and (ii) stopping CBT early.<br><br>Category (i) applies to both treatment arms, but category (ii) applies only to the medication plus CBT arm.<br><br>Therefore, a composite strategy will lead to the following different outcome definitions in each group:<br>-*medication plus CBT group*: being free from anxiety at 6 months without stopping either medication or CBT early<br>-*medication alone group*: being free from anxiety at 6 months without stopping medication early |

**Table 2 – Simulation study results**. Simulations are based on a trial comparing a 6-month vs. 12-month treatment duration on an "unfavourable" outcome, which is a composite of $Y_a$ (tuberculosis status at 12 months) and $Y_b$ (treatment discontinuation, i.e. whether patients in the 6-month arm stopped treatment before 6 months, or patients in the 12-month arm stopped treatment before 12 months). Estimated treatment effects are for the 6-month vs. 12-month comparison, meaning that negative treatment effects denote benefit for the 6-month treatment duration. In scenario 1, length of treatment has no direct effect on either $Y_a$ or $Y_b$, except through the longer time span in the 12-month arm which provides patients with more opportunity to discontinue. In scenario 2, the 6-month duration increases occurrences of $Y_a$ (i.e. is harmful), and has no direct effect on $Y_b$, except through the longer time span for the 12-month arm.

| Scenario | Performance measure | Estimated performance (MCSE) |
|---|---|---|
| 1: no difference on either $Y_a$ or $Y_b$ | Expected risk difference (6- vs. 12-month) | –10% (0.03) |
| | Rejection of the null | 89.2% (0.31) |
| | Excess events that could only occur in 12-month arm (out of 500 participants) | 49.9 (0.07) |
| 2: 6-month duration clinically harmful | Expected risk difference (6- vs. 12-month) | –1.7% (0.03) |
| | Rejection of the null | 8.1% (0.27) |
| | Excess events that could only occur in 12-month arm (out of 500 participants) | 50 (0.07) |

**Box 1 – Summary of key points.**

- Composite strategies to handle intercurrent events in the estimand definition are widely used.

- A potential benefit of the composite strategy over other options, such as hypothetical or principal stratum strategies, is that estimation is based on a randomised comparison, and so does not require untestable assumptions in order to be unbiased.

- However, a composite strategy when applied to intercurrent events such as treatment discontinuation or treatment switching can lead to a non-causal estimand.

- This occurs when the intercurrent event can be separated into distinct categories, such as based on time of occurrence, and at least one category does not apply to all treatment arms.

- In this situation, the use of a composite strategy results in the outcome being defined differently between treatment arms, which creates artificial differences in the comparison.

- A composite strategy should not be employed for intercurrent events which are related to changes in assigned treatment, unless it can be justified that this will not result in different outcome definitions between treatment arms.

**Box 2 – Recommendations when considering a composite estimand strategy.**

1. A composite strategy should not be employed when it results in outcomes being defined differently between treatment arms.

2. If a composite strategy is used for intercurrent events such as treatment discontinuation or use of rescue therapy, justification should be provided that demonstrates that the approach will not lead to different outcome definitions across treatment arms.

3. The estimand definition should clearly state that a composite strategy has been used (rather than simply implying this through the endpoint definition, which may not sufficiently alert readers to the potential issues).

4. When a composite strategy is used, each component should be defined as a secondary outcome.

**Funding**

BCK, TMP, CT, and TPM are funded by the UK MRC, grants MC_UU_00004/07 and MC_UU_00004/09.

**Contributions**

BCK wrote the first draft of the manuscript. TMP, CT, and TPM revised the manuscript. TPM and BCK developed the simulation study, and TPM implemented the simulation study. All authors read and approved the final manuscript.

**Declaration of Conflicting Interests**

The Authors declare that there is no conflict of interest.

**Acknowledgements**

The authors thank Jack Wilkinson for reviewing an early draft of the manuscript.